\documentclass[draft]{agujournal2019}
\usepackage{graphicx} 
\usepackage[inline]{trackchanges} 
\usepackage{amsmath,amsthm,amssymb}
\usepackage{natbib}
\usepackage{bm}
\usepackage{url}
\usepackage{multirow}
\usepackage{booktabs}
\usepackage{hyperref}
\draftfalse

\journalname{Geophysical Research Letters}

\begin{document}

\title{Reconstructions of Jupiter's magnetic field using physics-informed neural networks}

\authors{Philip W. Livermore\affil{1,4}, Leyuan Wu\affil{2,4}, Longwei Chen\affil{3,4}, Sjoerd de Ridder\affil{1}}

\affiliation{1}{School of Earth and Environment, University of Leeds, Leeds, LS2 9JT}
\affiliation{2}{Key Laboratory of Quantum Precision Measurement of Zhejiang Province, Center for Optics and Optoelectronics Research (COOR), Center for Optics and Optoelectronics Research (COOR), 
Collaborative Innovation Center for Information Technology in Biological and Medical Physics, 
College of Science, Zhejiang University of Technology, Hangzhou, 310023, China.}
\affiliation{3}{Guangxi Key Laboratory of Exploration For Hidden Metallic Ore Deposits, 
College of Earth Sciences, Guilin  University of Technology, Guilin, 541006, China.}
\affiliation{4}{Leeds Institute for Data Analytics, University of Leeds, Leeds, LS2 9JT}
\correspondingauthor{Phil Livermore; Leyuan Wu }{p.w.livermore@leeds.ac.uk; leyuanwu@zjut.edu.cn}

\begin{keypoints}
\item We present two reconstructions of Jupiter's magnetic field using physics-informed neural networks, based on orbits 1-33 and 1-50 of Juno.

\item Compared with spherical harmonic based methods, our reconstructions give clearer images at depth of Jupiter's internal magnetic field

\item Our models infer a dynamo at a fractional radius of 0.8.

\end{keypoints}

\begin{abstract}
Magnetic sounding using data collected from the Juno mission can be used to provide constraints on Jupiter's interior. However, inwards continuation of reconstructions assuming zero electrical conductivity and a representation in spherical harmonics are limited by the enhancement of noise at small scales. Here we describe new reconstructions of Jupiter's internal magnetic field based on physics-informed neural networks and either the first 33 (PINN33) or the first 50 (PINN50) of Juno's orbits. The method can resolve local structures, and allows for weak ambient electrical currents.  Our models are not hampered by noise amplification at depth, and offer a much clearer picture of the interior structure. We estimate that the dynamo boundary is at a fractional radius of 0.8. At this depth, the magnetic field is arranged into longitudinal bands, and strong local features such as the great blue spot appear to be rooted in neighbouring structures of oppositely signed flux.
\end{abstract}

\section*{Plain Language Summary}
A major goal of the Juno mission is to better constrain the interior structure of Jupiter. One method of doing this is to reconstruct Jupiter's magnetic field using measurements from Juno, which can then be used to find the dynamo region where the planetary magnetic field is generated. 
Standard assumptions of zero electrical conductivity and global solutions allow the reconstructions to be inwards extrapolated from where the data is collected on Juno's orbits, however this method of imaging is limited by amplified noise. Here, we present reconstructions based on recent advances in machine learning, in which the physical assumptions are relaxed and we allow for local structures. Our method shows a much clearer image of Jupiter's interior than has been possible before.

\section{Introduction}
The Juno mission, launched in 2011 \citep{Bolton_etal_2011}, has revolutionised our understanding of Jupiter's interior through the collection of both gravity and magnetic measurements in orbit since 2016. These new data have not only allowed new constraints on the density structure and zonal flow in the outermost parts of the planet \citep{Kaspi_etal_2018}, but have permitted new reconstructions of the magnetic field to unprecedented resolution \citep[e.g.][]{Connerney_etal_2017,Connerney_etal_2022}. These magnetic maps highlight local features such as the Great Blue Spot, sited within a largescale hemispheric field \citep{Moore_etal_2018} which shows evidence of secular variation \citep{Ridley_Holme_2016,Moore_etal_2019, Sharan_etal_2022,Bloxham_etal_2022,Connerney_etal_2022}.

In order to infer the structure of Jupiter's internally generated magnetic field, global reconstructions are needed that fit a physical model to the sparse magnetic dataset collected on orbital trajectories. 
The physical model commonly adopted is that the measured values come from a region free of electrical currents, and comprise signals dominated by the internally generated field with more minor contributions from an external magnetic field and unmodelled instrumentation noise. Typical studies then proceed by subtracting an approximation to the external field assuming a magnetodisk structure, with estimates of the parameters \citep{connerney1981modeling,Connerney_etal_2022}, although the difficulty in adopting an accurate representation is compounded by its unknown likely time-dependence \citep{Ridley_Holme_2016, Moore_etal_2019}. The remaining signal is then fit in a least-squares sense to an analytic description of an internally-generated magnetic field ${\bm B}$ using a potential $V$, with ${\bm B} = -{\boldsymbol \nabla} V$, which by construction exactly satisfies ${\bm J} = {\bm 0}$ where $\bm J = \mu_0^{-1} {\boldsymbol \nabla}\times {\bm B}$ is the ambient electrical current, where the constant $\mu_0$ is the permeability of free space. The potential is commonly represented in terms of a truncated spherical harmonic expansion \citep{connerney1981magnetic}, similar to comparable studies for Earth's magnetic field \citep[e.g.][]{alken2021international}.

Such reconstructions allow not only spatial interpolation between the Juno measurements, but also extrapolation into regions unconstrained by measurements. Radially inwards (i.e. downwards) continuation under Jupiter's surface, assuming the same electrically-insulating physics, is of particular interest because it allows inference of the dynamo radius, typical values for which are $0.8$ -- $0.83R_J$, where $R_J$ is Jupiter's equatorial radius, 71,492km  \citep{Connerney_etal_2022,Sharan_etal_2022}. However, this inwards continuation is numerically unstable because errors in small-scales, caused by leakage from unmodelled signals, become preferentially amplified compared with those of large-scale, eventually producing a signal swamped with noise.

In this paper, we propose a novel representation of Jupiter's internal magnetic field based on physics-informed neural networks (PINNs). Compared to standard approaches, our models give a similar reconstruction on and above Jupiter's surface but appear to be more stable under inwards continuation. 
In the following sections, we first describe the data before outlining our PINN approach. We present our new reconstructions and estimates of the dynamo radius, which we compare with those from existing methods, and end with a brief discussion. 

\section{Data}

\begin{figure}
    \centering
    \includegraphics[width=1\linewidth]{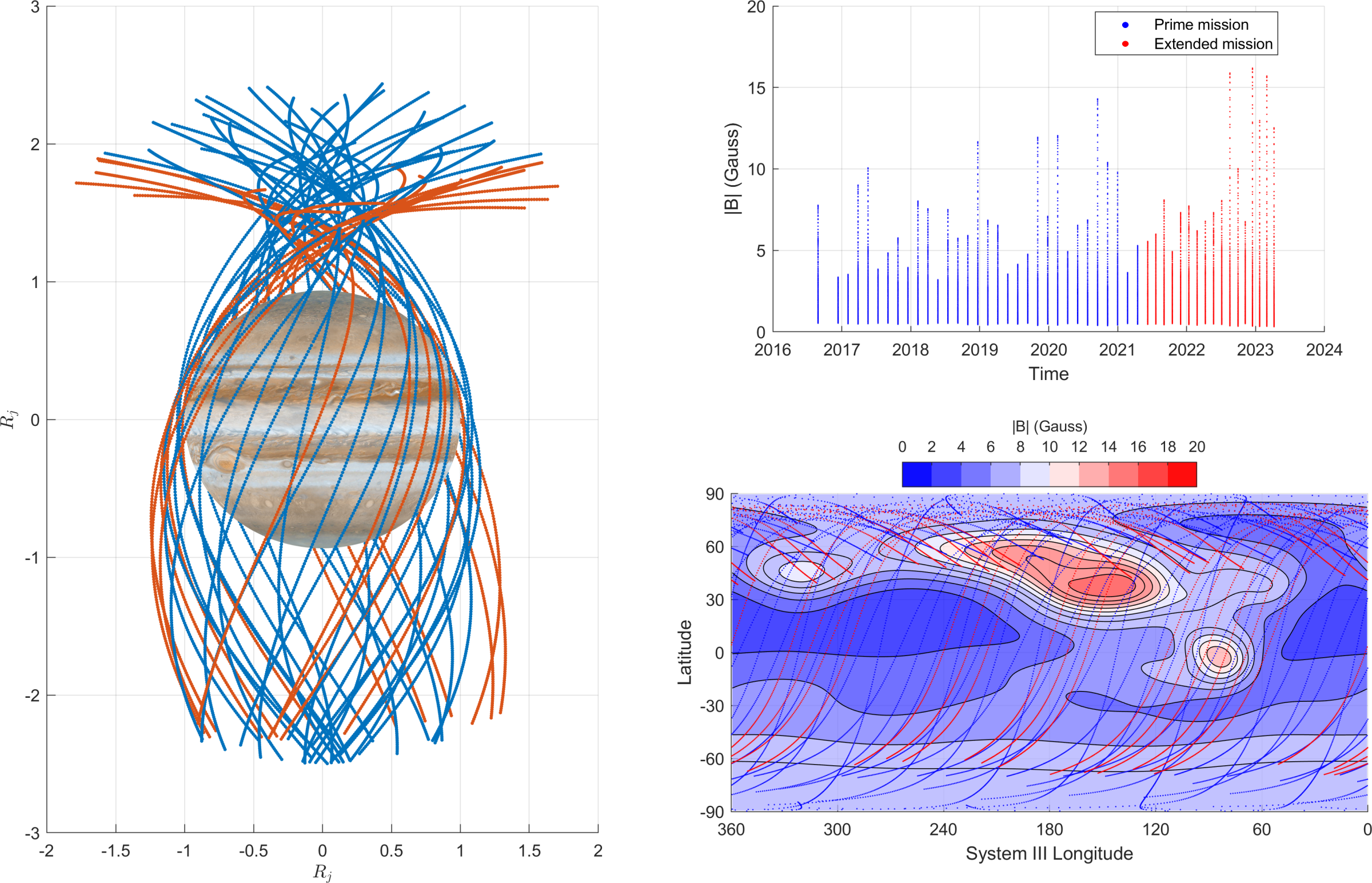}
    \caption{Juno data used in this work. Left: Juno's global coverage after 50 orbits, showing Juno's trajectory within radius $2.5~R_J$; the colours show the 33 prime mission orbits (red lines) and the extended mission, orbits 34 onwards (blue lines). Upper right: time span and magnitude range per orbit of Juno magnetic data. Lower right: orbital position (radius within $4.0~R_J$) projected onto a background contour map of the magnitude of magnetic field at $r=R_J$ reconstructed using model PINN50e.}
    \label{fig:fig1}
\end{figure}

Our work is based on vector magnetic field measurements made by Juno within its first $50$ perijoves during the period 2016 to 2023, which contains the prime mission of orbits 1-33. From these data we excluded the second perijove (PJ2) due to a spacecraft safe mode entry \citep{Connerney_etal_2018}.  The original observations were down-sampled to $30~s$ sampling rate, this being the approximate rotation time of the spacecraft, using a mean-value filter. 
At the small number of missing data points within each orbit, we average up to the missing data (even if less than 30 s) and restart our averaging after the missing data.
In order to maximise the internal signal content of the data, we used only measurements recorded at planetocentric spherical radius $r \le 4.0 R_J$. 
In total, there were $28011$ 3-component measurements of the magnetic field, of periapsis $1.02~R_J$ and taking magnitudes in the range of approximately $0.065-16$~Gauss. Figure \ref{fig:fig1} shows an overview of the data used in this work.

\section{Method}

Physics-informed neural networks, or PINNs, offer a technique for representing spatially dependent quantities by a neural network that are constrained not only by data but also physical laws \citep{Raissi2019}. 
We work in a planetocentric Cartesian coordinate system, and write the magnetic field in terms of a vector-potential: ${{\bm B}={\boldsymbol \nabla}\times{\bm A}}$, which satisfies the fundamental relation ${\boldsymbol \nabla} \cdot {\bm B} = 0$. The three independent components of $\bm A$, $(A_x, A_y, A_z)$, are expressed as individual feed-forward neural networks with $6$ hidden layers, $40$ neurons per layer and swish activation functions. These networks map position, rescaled to $[-1,1]^3$, to the potential.

We denote the set of tunable parameters (weights and biases) of the networks by $\bm \Theta$, and the representation of $\bm{A}$ and $\bm{B}$ as $\bm{A}_{\bm{\Theta}}(\bm{r})$ and $\bm{B}_{\bm{\Theta}}(\bm{r})$.
A physics-informed model is trained by minimizing the following loss function:
\begin{linenomath*}
\begin{equation}\label{eqn:obj}
     \mathcal{L}(\bm{\Theta}) = 
     w_{d}\mathcal{L}_{d}(\bm{\Theta}) + w_{p}\mathcal{L}_{p}(\bm{\Theta}),
\end{equation}
\end{linenomath*}
where
\begin{linenomath*}
\begin{equation}\label{eqn:obj1}
     \mathcal{L}_{d}(\bm{\Theta}) = \frac{1}{N_d} 
     \sum_{i}^{N_d} |\bm{B}_{\bm{\Theta}}(\bm{r}_{d}^{i}) - \bm{B}(\bm{r}_{d}^{i})|^2, \qquad 
    \mathcal{L}_{p}(\bm{\Theta}) = \frac{1}{N_p} 
    \sum_{i}^{N_p} |\left({\boldsymbol \nabla} \times {\bm B}_{\bm{\Theta}}\right)(\bm{r}_{p}^{i})|^2,
\end{equation}
\end{linenomath*}
are the data loss and physics loss terms with weights $w_d$ and $w_p$, $N_p$, $\bm{r}_{p}^{i}$ are the number and location of the collocation points used to constrain the physics loss, and $N_d$ are the number of Juno data used, each of which has location $\bm{r}_{d}^{i}$ and vector value $\bm{B}(\bm{r}_{d}^{i})$. The contribution to the data loss from each measurement is assumed equal, as is the contribution to the physics loss from each of the collocation points. With the optimisation so defined, both the physics and data loss terms can be quantified in physical units.
The quantities derived from $\bm{A}_{\bm{\Theta}}$, namely $\bm{B}_{\bm{\Theta}}(\bm{r})$ and ${\boldsymbol \nabla} \times {\bm B}_{\bm{\Theta}} = {\boldsymbol \nabla}({\boldsymbol \nabla} \cdot \bm{A}_{\bm{\Theta}}) - \nabla^2 \bm{A}_{\bm{\Theta}}$ are computed using automatic differentiation (AD) \citep{Baydin2018}.
All neural network models are built with the machine learning framework TensorFlow \citep{Abadi2016}, and trained with the built-in Adam optimizer \citep{Kingma2015} over 12,000 epochs with batch size 10,000. An empirical learning-rate annealing strategy, with an initial learning rate of $0.002$, and an exponential decay with a decay rate of $0.8$ and a decay step of $1,000$ iterations are adopted.
From tests with various network sizes, this network was just large enough to fit well all the data and physics constraints; a larger network with 7 layers instead of 6 did not change our results.  We do not use any explicit spatial regularisation in our method, although the curl in the physics loss penalises small-scales.

It remains to choose the weights and collocation points. Here we apply two techniques that improve the original choices of \cite{Raissi2019}. 
First, rather than prescribe the weight parameters $w_d$ and $w_p$, we allow them to be chosen dynamically. We fix $w_{p}=1$, but allow $w_d$ to change at each training epoch in order to balance the gradients of physical and data-fit loss with respect to the model parameters \citep{Wang2021}. 

Second, we adopt residual-based sampling for the physics loss term. 
While uniformly sampled collocation points for the physics-based term offers a simple approach, recent studies have shown promising improvements in  training accuracy by applying nonuniform adaptive sampling strategies \citep{Lu2021,Nabian2021,Wu2023}. Here we apply a simplified version of the residual-based adaptive distribution (RAD) method described in \cite{Wu2023}. For the first 3000 epochs we use a uniformly sampled set of points in a fixed region, but at epoch 3000 (and every 600 epochs thereafter) we create a probability density function, based on samples of the physics loss, which we use to resample the collaboration points, effectively increasing the local weighting in regions with a high physics loss.

It is important to highlight key differences between a PINN representation and existing reconstructions based on a spherical-harmonic potential. First, existing methods fit data in a weak sense (by least squares) to physics imposed in a strong form (by assuming an internal potential field representation, with exactly zero electric current density). This is quite different in a PINN, where both data and physics are fit in a weak form, which makes them particularly effective in problems when the data and physics are imperfectly known \citep{karniadakis2021physics}, as for Jupiter. Instead of assuming that $\bm J = 0$ and seeking a fit to an internally-generated magnetic field, instead we penalise the root-mean-squared electrical current density $\bm J$ which allows, for example, weak nonzero electric currents if the data require them. We therefore allow for some uncertainty in the current-free approximation.
Another key distinction is that we don't (and indeed cannot) separate internal and external fields as we fit the PINN to the fundamental physical law, rather than to an analytic solution which assumes the location of source. 
A third important difference is in the spatial representation. A spherical harmonic representation, an analytic solution to Laplace's equation, is defined by a set of Gauss coefficients, whose globally resolved wavelength is approximately $2\pi/(N+1/2)$, where $N$ is the maximum degree $N$ \citep{backus1996foundations}. In contrast, a neural network is a meshless method that can define both local and global solutions. It is defined by a set of weights and biases (here of the vector potential,  $\bm A$) that describe the internal coefficients of connected neurons, arranged in a structure that is governed by the number of neurons per layer, the number of layers, and the activation function. 

We create four time-independent PINN models, based on either the first 33 (PINN33i, PINN33e) or 50 Juno orbits (PINN50i, PINN50e). We deliberately distinguish between models internal to Jupiter (denoted by the character ``i'') which inwards continue into $r \le R_J$ the data observed in $r > R_J$, and those external to Jupiter (denoted by the character ``e'') which interpolate data within the same exterior region in which Juno measurements are made $r > R_J$. Models PINN50e, PINN33e were made first, using 300,000 collocation points within the region $1 \le r/R_J \le 4$. Models PINN50i and PINN33i were then constructed, using 40,000 collocation points within the region $0.8 \le r/R_J \le 1$; for these models the data loss term was replaced by a term describing matching in all three vector components to either PINN50e or PINN33e on $r=R_J$ at  80,000 randomly located points. 
Although mildly oblate, Jupiter is assumed spherical for simplicity.

\section{Results and discussion}

\begin{figure}
    \centering
    \includegraphics[width=1.0\linewidth]{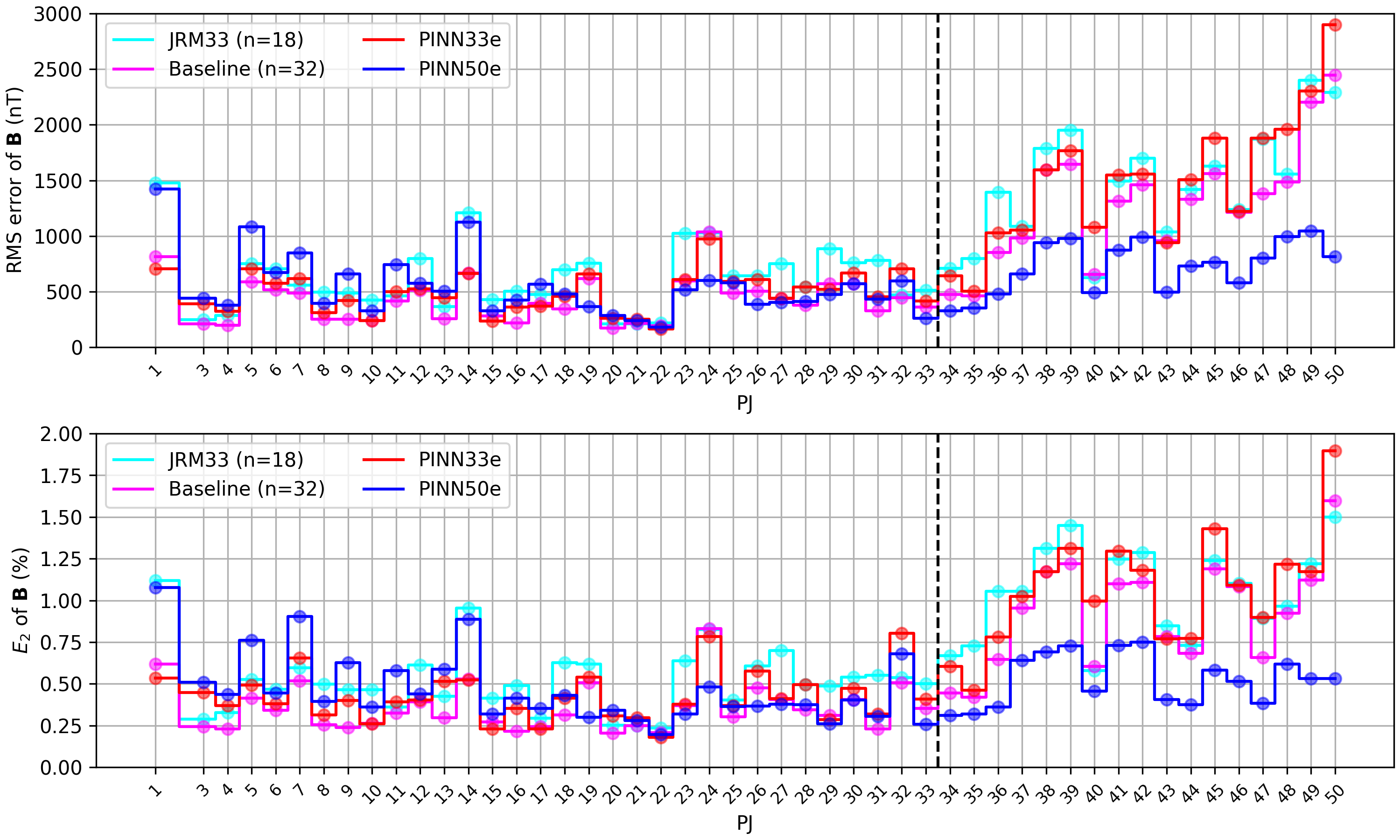}
    \caption{Orbital comparison of the discrepancy between various reconstructions of Jupiter's magnetic field: PINN33e, PINN50e, JRM33 and Baseline, with the Juno data. On each orbit, the error is quantified by taking the root mean squared value of the vector difference between the reconstructed magnetic field and the Juno measurements, similar to the data-loss term. We show the (upper) absolute value of this error, and (lower) relative value, $E_2$, of this error compared to the rms observed magnitude over the orbit. The dashed line delineates the prime from the extended mission. }
    \label{fig:orbital_errors}
\end{figure}

Figure \ref{fig:orbital_errors} shows the data loss, in terms of an orbital comparison of the difference between Juno data and four models: PINN33e,  PINN50e and two recent spherical harmonic models JRM33 ($N=18$) \citep{Connerney_etal_2022} and the Baseline model of \citet{Bloxham_etal_2022} with $N=32$.
These recent models have been chosen because although they are both based on the first 33 orbits, they differ in how the spherical harmonics are fitted: JRM33 uses an approach based on singular value decomposition, whereas the Baseline model uses regularisation. Both of these studies adopt a magnetodisk approximation to the external field \citep{Connerney_etal_2022} which we include alongside the spherical harmonic representation of the internal field; the PINN models represent both internal and external field. 

The models based only on the prime orbits (1-33, excluding 2): PINN33e, JRM33 and Baseline show a comparable absolute rms error per orbit. For the majority of orbits, PINN33e has an error less than JRM33, with a few exceptions such as orbits 3, 32. Over the first 33 orbits as a whole, the rms error for JRM33 is 680.5 nT, compared with 465.0 nT for Baseline and 519.7 nT for PINN33e.
Applied to the data from orbits 34-50, these models exhibit a discrepancy with the measurements which grows with time, providing additional evidence for Jupiter's secular variation. Model PINN50e has a slightly higher rms error of 599.2 nT for orbits 1-33, but fits the data for orbits 34-50 much better because it has been trained in part on these data.

\begin{figure}
    \centering
\includegraphics[width=0.95\linewidth]{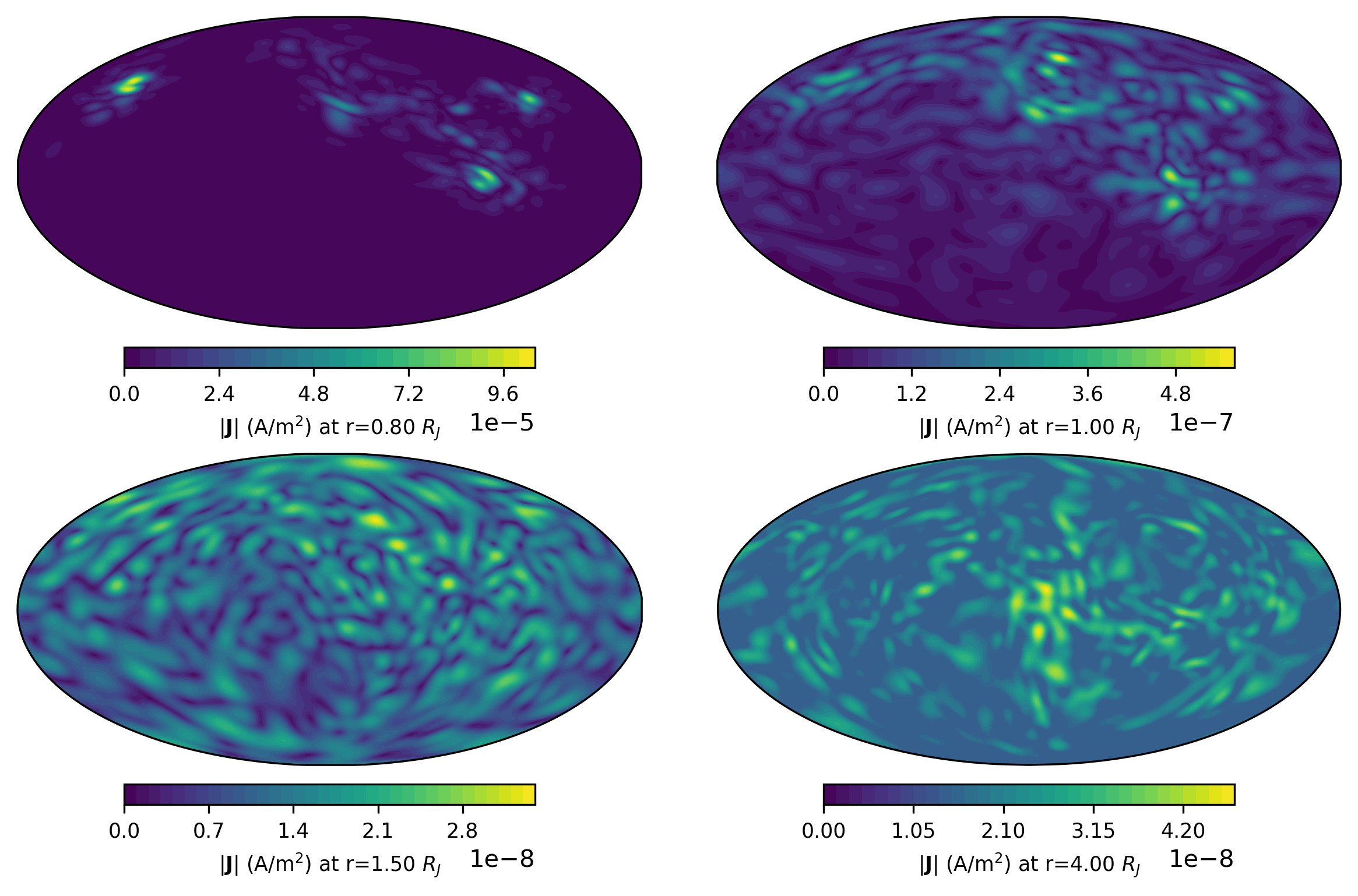}
    \caption{The magnitude of current density $|\bf J|$ from PINN50e and PINN50i shown on illustrative radii $(r/R_J = 0.8, 1, 1.5, 4$) on a Mollweide projection with the central meridian at a longitude of $180^o$ west (System III coordinates).}
    \label{fig:J}
\end{figure}

Figure \ref{fig:J} shows the physics-loss by contours of the magnitude of electrical current density $|\bf J|$ on selected radii. For radii $r>R_J$, the magnitude of the current density of the PINN50e model is about $10^{-8}$A/m$^2$, increasing with decreasing radius to about $10^{-6}$A/m$^2$ at $r=0.8 R_J$. 
The current density includes a signature from not only any external electrical currents, but also electro-magnetic structures which are numerically favoured because they allow a good fit to the data. Estimates of the current density associated with a simple magnetodisk model are about $10^{-9}$A/m$^2$ at $r=5R_J$ \citep{connerney1981magnetic}, which is consistent with our values in $r>R_J$. The increase of $|{\bf J}|$ by 100 from $r=4$ to $r=0.8$ is principally explained by the increase in magnetic field strength by a similar factor due to proximity to the dynamo source. Structurally, at large radii, $|\bf J|$ appears dominated by weak small-scale numerical artefacts; however, as the radius decreases, $|\bf J|$ becomes dominated by gradients in $\bm{B}$, focussed at locations where the magnetic field is largest.

\begin{figure}
    \centering
    \includegraphics[width=1.0\linewidth]{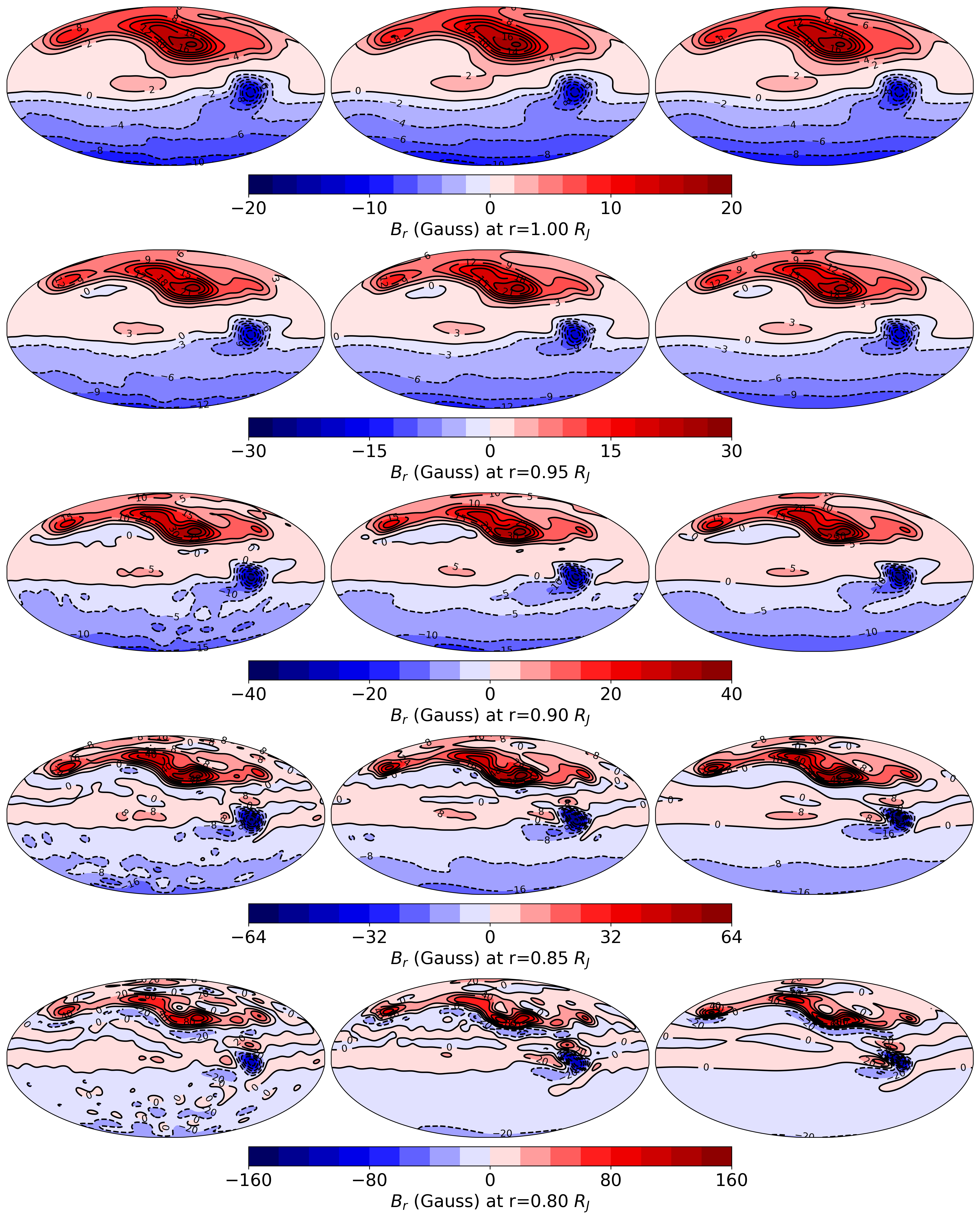}
    \caption{The radial component of Jupiter's magnetic field on various spherical radii inside Jupiter's surface. The plots are shown on a Mollweide projection with the central meridian at a longitude of $180^o$ west (System III coordinates). Left column shows the JRM33 model, $N=18$ \citep{Connerney_etal_2022}, the middle column shows the Baseline model, $N=32$  \citep{Bloxham_etal_2022} and the right column shows the model PINN50i. }
    \label{fig:Br}
\end{figure}

The structure of JRM33, Baseline and PINN50i at radii $r/R_J = 1, 0.95, 0.9, 0.85, 0.8$ are shown by contours of radial field in figure \ref{fig:Br}. 
On $r=R_J$ the models are almost indistinguishable in terms of physical structure, but as the radius decreases, the magnetic field strength increases and the lengthscales decrease. The instability of inwards continuation in the spherical harmonic models is readily apparent by the prevalent fine-scaled noise, particularly in the azimuthal direction. The JRM33 model in particular has a lot of small-scaled structure in the southern hemisphere, which appears to be noise as it is entirely absent in the other smoother models. Of the three models, PINN50i is smoothest and remains relatively free of longitudinal small-scales; consequently the features at depth are much easier to identify.  

At $r \le 0.85 R_J$, the field appears arranged into longitudinal bands of positive flux, with a strong band at high latitude and weaker bands near the equator. Many of the strong patches of flux have adjacent oppositely signed counterparts, as can be seen in particular around the root of the great blue spot. The hemispheric structure is also striking, with almost all the magnetic structure of the field being confined north of the equator.

\begin{figure}
    \centering
    \includegraphics[width=1.0\linewidth]{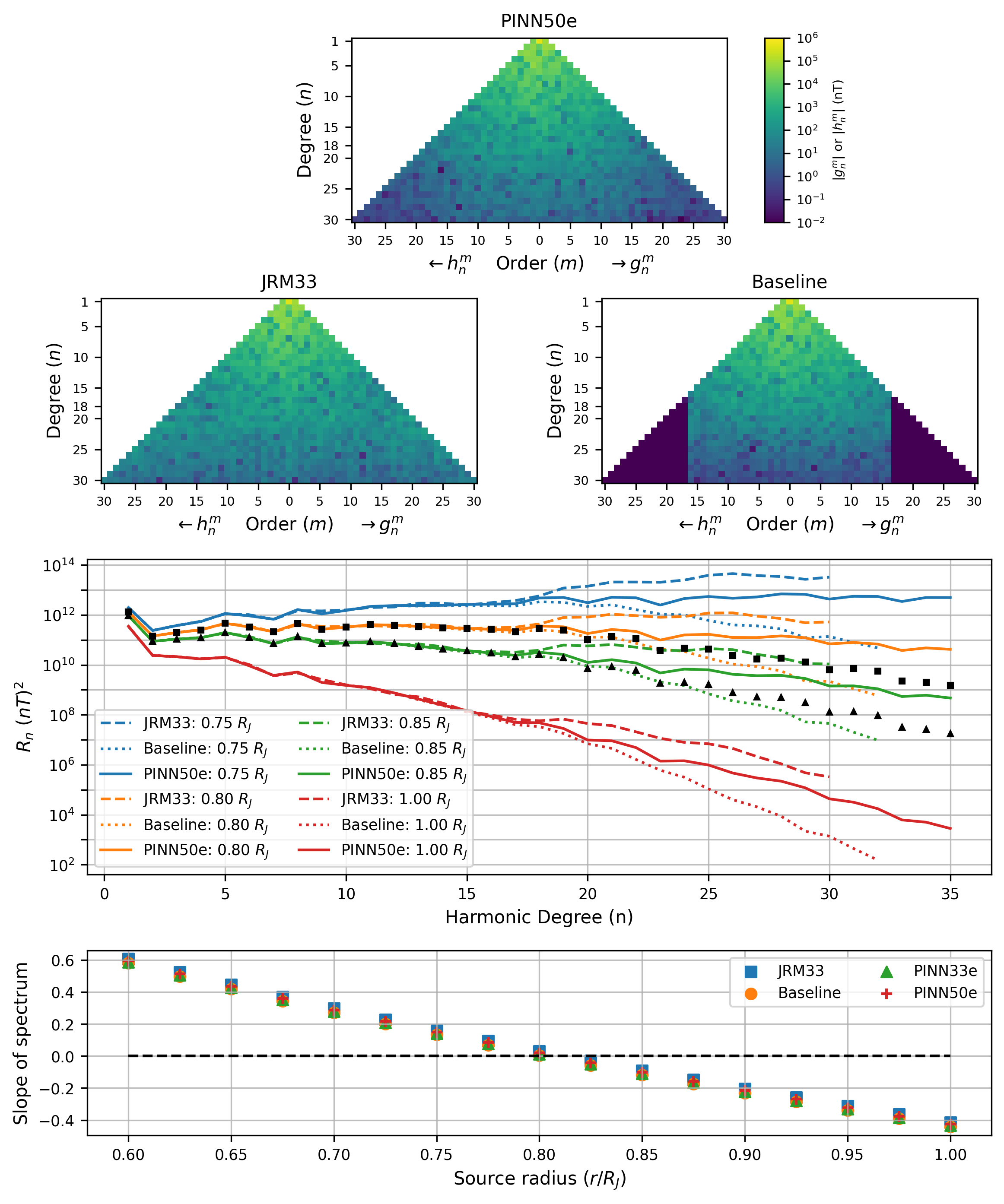}
    \caption{Upper panel: Gauss coefficients of PINN50e at $r=R_J$, JRM33 and Baseline. Middle panel: Lowes-Mauersberger spectrum of three inwards analytically continued models (coloured lines): PINN50e ($N=35$), JRM33 ($N=30$) and Baseline ($N=32$); black symbols show similar spectra obtained from extrapolation using PINN50i in $r<R_J$
    (squares: $0.80 R_J$; triangles: $0.85 R_J$). Lower panel: spectral slope (fit to degrees 2--18), with radius for four analytically continuated models: JRM33, Baseline, PINN33e and PINN50e.}
    \label{fig:spec}
\end{figure}

A common approach to determining the dynamo radius is by identifying where the Lowes-Mauersberger spectrum of the magnetic field  \citep{lowes1974spatial,mauersberger1956mittel} is flat, which describes a white-noise source. This procedure relies on the spherical harmonic representation of the magnetic field:
\begin{linenomath*}
\begin{equation}
     {\bm B} = -R_J\, {\boldsymbol \nabla} \ \sum_{n=0}^N \sum_{m=0}^n \left( \frac{R_J}{r}\right)^{n+1} \left[ g_n^m P_n^m(\theta) \cos(m\phi) + h_n^m P_n^m(\cos\theta) \sin(m\phi) \right]
\label{eq:SH_rep}
\end{equation}
\end{linenomath*}
and $g_n^m$ and $h_n^m$ are the Gauss coefficients of degree $n$ and order $m$ and $P_n^m$ are associated Legendre functions.  The spectrum is then derived as
\begin{linenomath*}
\begin{equation}
R_n = (n+1)\left(\frac{R_J}{r}\right)^{(2n+4)}
\sum_{m=0}^n(g_n^m)^2 + (h_n^m)^2.
\label{eq:LM_spec}
\end{equation}
\end{linenomath*}
whose profile with $n$ depends on the radius $r$. 
The upper part of figure \ref{fig:spec} shows a comparison of the Gauss coefficients for PINN50e (projected onto \eqref{eq:SH_rep} at $r=R_J$), JRM33 and Baseline. They are all very similar at large-scales (small $n$, $m$), in accordance with the agreement at $r=R_J$ (figure \ref{fig:Br}). The spherical harmonic model JRM33 was constructed with no regularisation, and has higher power at small scales ($n,m \ge 20$) than the other two models. The PINN50e model has no explicit regularisation either, but small-scale features are inherently penalised through the spatial curl in the physics loss term. As a consequence, in PINN50e there is a noteable decrease in power at small scales $n,m \ge 20$, somewhat mirroring the result of the explicit damping applied in the Baseline model.

For the PINN models, in order to find the radius where the spectrum is flat, we have two options. First is analytic continuation, where we project the field at $r=R_J$ onto \eqref{eq:SH_rep} and use the inherent radial dependence within \eqref{eq:LM_spec}. 
Second, we can use PINN extrapolation, for which we use PINN50i to inwards continue, and at each radius $r<R_J$, project onto \eqref{eq:SH_rep} and then use \eqref{eq:LM_spec}. Both procedures remove any external field within the PINN model. In either case, we find the Gauss coefficients by performing a spherical harmonic transform of the spherically radial component $B_r$.

The middle panel of figure \ref{fig:spec} shows the Lowes-Mauersberger spectrum as a function of degree $n$ for JRM33, Baseline and PINN50e (coloured lines: analytic continuation; black symbols: PINN extrapolation). At $r=R_J$ the spectral power for degrees 2--18 agrees well between the models and falls off exponentially with $n$. The power in the dipole is higher than a simple profile predicts. 
As the radius is decreased the profile flattens as the smaller scales become more prominent. Above degree 18, the three analytically continued models diverge, with JRM33 having the most power at high degree. Of the three models, the Baseline model (which is the only model with explicit regularisation) has the least power at small-degree.
Comparing the analytic and PINN extrapolation methods, although they agree on $r=R_J$ by construction, for $r < R_J$ and degrees higher than about 18 they diverge, with the PINN extrapolation having smaller power at high-degree.  

We quantify the slope of the spectrum by fitting a straight line to $\log_{10} R_n(n)$ for degrees 2--18.  The lower panel of figure \ref{fig:spec} shows the slope variation with radius for four models analytically inwards continued using \eqref{eq:SH_rep}; the extrapolated PINN models give very similar results.
On making the assumption that the slope is zero at the source we infer that the dynamo radius is about $r=0.8R_J$, in approximate agreement with other studies \citep[e.g.][]{Connerney_etal_2022, Sharan_etal_2022}.

\section{Concluding remarks}
We have presented a reconstruction of Jupiter's magnetic field, based on data from Juno within the framework of a physics-informed neural network.
Our reconstructions have a similar misfit to to the data compared with other spherical harmonic methods, and produce a similar structure of magnetic field on Jupiter's surface.
However, by using a meshless method, and only weakly constraining the (poorly known) physics, our models are not apparently hostage to the typically enhanced noise with decreasing radius. Compared with spherical harmonic-based methods, we produce a clearer image at depth of the localised interior magnetic field.

The fact that most of the structure in Jupiter's field appears confined to the northern hemisphere perhaps makes neural networks a particularly effective modelling tool. Even at modest resolution, neural networks are able to very well represent local structures, compared to spherical harmonics which are inherently global. 
Future applications of the PINN method include quantifying the secular changes close to Jupiter's dynamo region, and applications to other planets.

\section*{Data Availability Statement}
The original Juno magnetometer data are publicly available on NASA's Planetary Data System (PDS) at Planetary Plasma Interactions (PPI) node at \url{https://pds-ppi.igpp.ucla.edu/search/?sc=Juno\&t=Jupiter\&i=FGM}. The produced PINN models, together with input processed Juno data, spherical harmonic models, and all related Python code and Jupyter notebook to reproduce all the results in this work, are archived in the Github repository \url{https://github.com/LeyuanWu/JunoMag\_PINN\_VP3}.

\acknowledgments
This study was funded by the National Natural Science Foundation of China (Grant No. 42374173), National Natural Science Foundation of Guangxi Province of China (Grant No. 2020GXNSFDA238021).  This work was undertaken on ARC4, part of the High Performance Computing facilities at the University of Leeds, UK. We thank Jack Connerney for help with accessing the Juno data.
\newpage
\bibliography{Refs}

\end{document}